# Understanding spin textures in *PT*-broken systems through universal symmetry-constrained rules

Wei Tan[1,4], Jianfeng Wang[2,4], Yang Li[1], and Bing Huang[1,3,5,*]

[1]*Beijing Computational Science Research Center, Beijing 100193, China*
[2]*School of Physics, Beihang University, Beijing 100191, China*
[3]*School of Physics, Beijing Normal University, Beijing 100875, China*
[4] These authors contributed equally
[5] Lead contact
*Correspondence: bing.huang@csrc.ac.cn

## SUMMARY

**The current understanding of spin-polarization phenomena in crystals relies heavily on the development of specific *k·p* Hamiltonians. A more fundamental and symmetry-driven understanding, based solely on crystalline symmetries, remains largely incomplete. In this work, we develop a universal theory consisting of five symmetry-constrained rules to capture the key framework of diverse spin textures (STs) in *PT*-broken systems. These rules comprehensively classify spatial operations and their combinations with anti-unitary operation *T* in symmetry-preserved *k*-invariant subspaces. The theory allows us to address puzzling STs and facilitates the ab initio design of unconventional STs in spin-orbit coupled crystals. Key examples include the identification of vertex-like and windmill-like STs in materials with $D_n$ and $C_{nv}$ point groups, extending beyond the currently known Radial and Rashba STs. By integrating different types of antiferromagnetic configurations, we also achieve radial STs and the Dresselhaus effect in centrosymmetric systems, challenging the long-standing assumption that these effects exist solely in non-centrosymmetric systems. When combined with topological band theory, we further propose that symmetry-constrained three-dimensional persistent STs can widely exist in two types of non-symmorphic crystals, effectively resolving the ongoing debate regarding their existence.**

## KEYWORDS

**Symmetry constrained rules, PT-broken crystals, spin polarization effect, vertex-like spin textures, windmill-like spin textures, 3D persistent spin textures.**

## INTRODUCTION

Symmetry is the foundation of the universe and a fundamental framework for understanding and predicting a wide range of physical phenomena in condensed matter. For example, in crystals with *PT* symmetry, the presence of spin-orbital coupling (SOC) can induce the degeneracy of energy bands, leading to hidden spin polarization [1-4]. When *PT* symmetry is broken, the SOC can lift this degeneracy, resulting in nonzero spin polarization [5]. Importantly, many intriguing spin-related phenomena, such as the Edelstein effect [6-10], spin-orbit torque [11-

[14], spin galvanic effects [15-17], and spin relaxation [18-24] are attribute to unique spin textures (STs). These textures consist of the spin-polarization vectors of different ***k*** points in Brillouin zone. The STs reveal the inherent interplay between charge and spin degrees of freedom under the influence of the crystal field.

Historically, the most well-known STs are the Rashba and Dresselhaus ones, which were initially discovered in wurtzite [5,25,26] and zinc-blende [5,27] crystals more than half a century ago. Recently, the emergence of unexpectedly new STs, *e.g.*, radial STs in chiral crystal, such as elemental Te [28-30], and two-dimensional (2D) persistent STs (PSTs) in non-centrosymmetric systems with point groups $C_{3h}$ and $D_{3h}$ (such as $Ge_3Pb_5O_{11}$ [31]), has sparked intensive interest in exploring their physical origins [32-36]. Point-group-dependent low-order effective Hamiltonians have been successfully developed to understand these specific STs [5,25-28,31,34-36]. However, a universal theory that relies solely on symmetry arguments to simply and quickly capture the key framework of different STs in SOC systems is still lacking.

In this article, we present a universal theory, consisting of five basic symmetry-constrained rules, to understand the highly variable STs in $PT$-broken and SOC crystals. We systematically enumerate all the spatial operations and their combinations with anti-unitary time-reversal operation $T$ to derive these rules. Our theory, validated by extensive examples using first-principles density-functional theory (DFT) calculations (see **METHODS**), can be utilized to explain existing puzzling STs or to design novel STs for spintronic applications. First, we demonstrate the existence of two types of zero spin-polarization lines (ZSPLs) in crystals. Next, by incorporating the magnetic point group (MPG) symmetry, we find that vertex-like and windmill-like shapes of STs can exist in materials with $D_n$ and $C_{nv}$ point groups ($n$=3, 4 and 6), respectively, in addition to the previously known Radial and Rashba STs. Furthermore, by combining real-space antiferromagnetic (AFM) orders, we show that radial STs and the Dresselhaus effect could even emerge in centrosymmetric crystals. Finally, we propose symmetry-constrained 3D PSTs may be abundant in two families of non-symmorphic crystals. We emphasize that our rules can also be applied to understand diverse orbital textures in crystals, as the orbital angular momentum and the spin-polarization vector share identical symmetry transformation properties.

## RESULTS

### Complete Rules for Spin-polarization

Considering a general symmetry operation $O$, it keeps certain wave-vectors $k$ invariant. These invariant vectors constitute a subspace $M[O]$ of the whole Brillouin zone ($M[O] = \{\boldsymbol{k}|O\boldsymbol{k} = \boldsymbol{k}\}$) [37]. The observable spin-polarization vector ***S*(*k*)** is severely constrained by $O$ on $M[O]$ and typically corresponds to a nondegenerate state; otherwise, one must sum over the degenerate states [1].

Firstly, we demonstrate that symmetry operations in the SOC system depends solely on four basic operations: inversion $P$, time-reversal $T$, integer/fractional lattice translation $\tau$, and rotation $C_n$, along with their combinations (see **METHODS**). Secondly, we explore all symmetry constraints on ***S*(*k*)**. The presence of SOC breaks the spinor operation $U$ of SU(2) group, completely locking spatial and spin rotations [38,39]. The sufficient and necessary condition for nonzero spin-polarization across the entire Brillouin zone is $PT$ breaking (In **Table S3**, we enumerate 90 *PT*-broken MPGs and their symmetry operations). The latter guarantees *n*-fold rotation $C_{nz}$ satisfying $C_{nz}\boldsymbol{S}(\boldsymbol{k}) = (C_{nz}S_x, C_{nz}S_y, S_z)(C_{nz}k_x, C_{nz}k_y, k_z)$, which constitutes the physical origin of diverse ST in *PT*-broken crystals. Finally, after a complete symmetry analysis (see **METHODS**), we can obtain five basic rules describing the symmetry constraints on spin-polarization:

(i) *$C_n$-constrained Parallel Spin-polarization*: Only parallel components $S_z(k)$ are nonzero in *$M[C_{nz}]$* (i.e., rotation axis) [**Figure 1A, *Rule I***].

(ii) *m-constrained Out-of-plane Spin-polarization*: Only out-of-plane components $S_y(k)$ are allowed in *$M[m_y]$* (i.e., mirror plane) [**Figure 1B, *Rule II***].

(iii) *$\bar{n} \times T$-constrained ZSPL*: A null spin polarization emerges along *$M[\bar{n} \times T]$* (i.e., $\bar{n} \times T$-invariant axis) [**Figure 1C, *Rule III***].

(iv) *$C_2 \times T$-constrained In-plane Spin-polarization*: Only in-plane components $S_x(k)$ and $S_y(k)$ are nonzero in *$M[C_{2z} \times T]$* (i.e., $C_{2z} \times T$-invariant plane) [**Figure 1D, *Rule IV***].

(v) *$m \times T$-constrained Perpendicular Spin-polarization*: A null component $S_y(k)$ emerges along *$M[m_y \times T]$* (i.e., $m_y \times T$-invariant plane) [**Figure 1E, *Rule V***].

Note that the symbol $\tau$ has been omitted for convenience. Since $\tau\boldsymbol{S}(\boldsymbol{k}) = \boldsymbol{S}(\boldsymbol{k})$, our rules also apply to non-symmorphic symmetries. Based on ***Rules I-V***, we see that ***S*(*k*)** is always parallel to the rotation axis and $C_2 \times T$-invariant plane, perpendicular to the mirror plane and $m \times T$-invariant axis, or exhibits a ZSPL along $\bar{n} \times T$-invariant axis in symmetry-preserved *k*-invariant regions. We have conducted extensive material examples to confirm the universality of our rules (**Figures S1-S15**).

In principle, these rules or their combinations can be applied to understand the kaleidoscopic STs (or orbital textures) in *PT*-broken crystals, particularly around a high-symmetry point, even without knowledge of their specific Hamiltonians. First, one must determine the MPG symmetry G(***k****) of the corresponding high-symmetry point ***k****. ***S*(*k**)** usually corresponds to a null magnitude when G(***k****) includes various symmetry operations (for example, in a *422.1* MPG, where ***S*(*k**)** is aligned with the *x* and *z* axes, resulting in a null magnitude). Next, preliminary frameworks of ST centered at ***k**** point can be constructed based on our rules and G(***k****). Third, the spin-polarization distribution deviating from high-symmetry lines or planes can be viewed as the continuous evolution of ***S*(*k*)** between adjacent high-symmetry lines or planes, assuming that ***S*(*k*)** changes continuously as a function of perturbations with the same irreducible representation[33]. This continuity assumption is generally valid within a certain range of *k* vectors unless there a perturbation is strong enough to break this continuity.

### *Classification of zero spin polarization lines*

While ***Rule III*** naturally indicates the existence of ZSPLs in *PT*-broken systems, it is important to provide a comprehensive understanding of ZSPLs, a puzzling phenomenon that has been accidentally observed in many systems [31,40,41]. Interestingly, we find that the ZSPLs can be classified into two types. Type-I ZSPLs are pinned, which include two major cases: (i) According to ***Rule III***, there is a pinned ZSPL along $\bar{n} \times T$-invariant axis; (ii) By combining ***Rules I*** and ***II***, when two mirror symmetries are included and related by rotation (*e.g.*, $m_x$ and $m_y$, and $C_{2z}$=$m_x \times m_y$), a ZSPL is formed and pinned along the intersecting lines of $m_x$ and $m_y$, corresponding to the $C_{2z}$ axis [40]. Type-II ZSPLs are unpinned, which are band-dependent and depend on the presence of both mirror operations and the anti-unitary operation *T*. Since only out-of-plane spin components are permitted in the mirror plane (***Rule II***), the presence of *T* flips ***S***(***k***) from +*k* point to -*k* point in the plane, *i.e.*, $\boldsymbol{S}(\boldsymbol{k}) = -\boldsymbol{S}(-\boldsymbol{k})$, creating at least one unpinned boundary line with zero spin-polarization that separates regions with opposite perpendicular ***S***(***k***). In addition, Type-II ZSPLs can exist in magnetic systems that possess $T\tau$ and single *m* operations, corresponding to Type-IV magnetic space groups (MSGs).

To confirm our classification, we selected nonmagnetic crystals with $C_s$ and $S_4$ point groups. The first example is $ReN_2$. As shown in **Figure 2A**, there is a single mirror operation $m_{010}$ and an anti-unitary operation *T*. In **Figure 2B**, DFT-calculations reveal that only out-of-plane $S_y$(***k***) is nonzero in $k_y$=0 plane (mirror plane), confirming ***Rule II*** and forming 2D PSTs. Notably, there exists one type-II ZSPL in the single mirror plane [green-line in **Figure 2B**], separating regions with opposite perpendicular ***S***(***k***). The position of the ZSPL varies for different energy bands (**Figure S1**). In addition, the parallel component $S_y$(***k***) is forbidden along the $k_y$ axis ($m \times T$-invariant axis), confirming ***Rule V***. The second example is $GeSe_2$. As shown in **Figure 2C**, there is a roto-inversion $\bar{4}_{010}$ and a twofold-rotation $C_{2,010}$ along the *y* axis. Using DFT calculations, we observe that Type-I ZSPL appears along the $k_y$ axis ($\bar{4}_{010} \times T$-invariant axis), confirming ***Rule III***. Unlike **Figure 2B**, the position of the ZSPL is fixed for different bands (**Figure S2**). Furthermore, only in-plane components $S_x$(***k***) and $S_z$(***k***) are nonzero in the $k_y$=0 plane ($C_2 \times T$-invariant plane) [**Figure 2D**], confirming ***Rule IV***.

### *Two types of spin frameworks with $D_n$ and $C_{nv}$ point groups*

Our rules can be utilized to construct the key spin frameworks corresponding to MPG symmetry. Interestingly, we find that two types of spin frameworks can exist in materials with $D_n$ and $C_{nv}$ point groups (*n*=3, 4 and 6). This challenges the widely-accepted understanding that materials with $D_n$ and $C_{nv}$ point groups typically correspond to radial-like and Rashba-like STs, [29,30,32,36] thereby broadening the current understanding of STs in crystals.

Next, we will explain the physical origin of the two types of spin frameworks by applying ***Rules I-II*** and ***IV-V***. Firstly, we consider the purely $D_n$ point group as an example [**Figure 3A-3F**]. There are *n* in-plane two-fold rotations that lie in the $k_z$=0 plane. According to ***Rule I***, strict parallel-spin-momentum locking exists along the rotation axes, where ***S***(***k***) is either parallel or anti-parallel to their momentum ***k***. However, we find that all two-fold rotations can be classified into two types of nonequivalent rotation axes. For instance, in the $D_3$ point group, these can be classified into $2_{100}$ ($2_{010}$/ $2_{110}$) and $2_{-100}$ ($2_{0-10}$/ $2_{-1-10}$) axes. ***S*(k)** can exhibit different magnitudes and orientations in the nonequivalent rotation axes. Therefore, we can predict two types of preliminary frameworks: (i) For **type-I frameworks** with the purely $D_n$ point group [**Figure 3A-3C**], the main feature is captured by considering two types of ***S***(***k***) along the rotation axes [***Rule I***], which are uniformly oriented parallel or antiparallel to ***k***, but may have different magnitude, resulting in an in-plane radial-like shape of spin textures. (ii) For **type-II frameworks** with the purely $D_n$ point group [**Figure 3D-3F**], the main features arise from two types of ***S*(k)** along the rotation axes, with directions oriented parallel and antiparallel to their momentum *k*, leading to an in-plane vertex-like shape of spin textures. Additionally, the absence of $C_{2,001} \times T$ symmetry could result in nonzero out-of-plane components in the zone deviating from the rotation axes, where the spin distribution can be viewed as consisting of *2n* parts due to the presence of *n* in-plane two-fold rotations [**Figure 3A-3F**].

When considering the $D_n$ point group with $T$ symmetry, two notable differences emerge: (i) For the spin vector field with the $D_3$ point group, the main feature can be characterized by two types of ***S*(*k*)** along the rotation axes and vertical $C_2\times T$ planes, according to ***Rule I*** and ***IV***, respectively. In this case, the $2_{100}$ ($2_{010}$/ $2_{110}$) axes become equivalent to $2_{-100}$ ($2_{0-10}$/ $2_{-1-10}$) axes due to the presence of $T$ symmetry [**Figure 3G and 3J**]. (ii) For the spin vector field with the $D_4$/$D_6$ point group, ***S*(k)** exhibits a null out-of-plane component due to the presence of $C_{2,001}\times T$ symmetry [**Figure 3H-3I and 3K-3L**].

A similar analysis can be applied to systems with the $C_{nv}$ point group: (i) For **type-I frameworks** with purely $C_{nv}$ point groups [**Figure 3M-3O**], the main feature can be captured by considering two types of ***S*(*k*)** in the mirror planes [***Rule II***], which are uniformly oriented either clockwise or anticlockwise to ***k***, but with possibly different magnitude. This results in an in-plane Rashba-like shape of spin textures. (ii) For **type-II frameworks** with purely $C_{nv}$ point group [**Figure 3P-3R**], the main features can be captured by considering two types of ***S*(k)** in the mirror planes, with directions oriented clockwise or anticlockwise to their momentum *k*, respectively, leading to an in-plane windmill-like shape of spin textures. When considering the additional $T$ symmetry [**Figure 3S-3X**], two similar differences can be observed as those in **Figure 3G-3L**.

Interestingly, at the same high-symmetry point, we discover that the emergence of two different spin-texture frameworks primarily depends on the specific constant-energy contour at different energy bands. To demonstrate our classification of these two spin frameworks, we select the $D_4$ and $C_{4v}$ point groups in real materials. For the $D_4$ point group, we examine the nonmagnetic $TeO_2$ (space group $P4_32_12$). DFT calculations confirm that a radial spin texture appears in the bands with an isotropic constant-energy contour [**Figure 4A**, type-I case of **Figure 3H**], while a vertex-like spin texture manifests in the bands with an anisotropic constant-energy contour [**Figure 4B**, type-II case of **Figure 3K**]. For the $C_{4v}$ point group, we select the nonmagnetic $AlN_2OTa$ (space group *P4mm*). Notably, a Rashba spin texture is observed in the bands with an isotropic constant-energy contour [**Figure 4C**, type-I case of **Figure 3T**], whereas a windmill-like spin texture exists solely in the bands with an anisotropic constant-energy contour [**Figure 4D**, type-II case of **Figure 3W**]. Importantly, these spin textures exhibit null out-of-plane components, as predicted in **Figure 3**.

### ***Unconventional spin-polarization effect in centrosymmetric crystals***

The natural ideal is to explore whether nontrivial STs could exist in conventional magnetic materials. Here, we demonstrate for the first time that radial STs can be generated in a centrosymmetric system with the aid of specific magnetic order, based on our proposed rules. This challenges the widely accepted notion that radial spin textures can only occur in chiral materials[9,29,30,34,36]. Additionally, we explore the physical origin of Dresselhaus-like STs in the noncollinear AFM systems with MPG $\bar{3}m'$. These two cases broaden the scope for designing spintronic devices from the perspective of magnetic symmetry rather than solely crystalline symmetry.

For the first case, we select conventional AFM $ZnV_2O_4$ to confirm the presence of radial STs with parallel spin-momentum locking. The centrosymmetric $ZnV_2O_4$ features a collinear AFM-order ground-state (MSG $P_I4_32_12$, MPG $422.1'$) [42], where adjacent magnetic V atoms have opposite spin orientations, breaking $PT$ symmetry, as shown in **Figure 5A. Figure 5B** displays their first Brillouin zone. The symmetry operations at the $\Gamma$-point include {$E$, $\tilde{T}$, $C_{4z}$, $C_{2z}$, $C_{2x}$, $C_{2y}$, $C_{2xy}$, $C_{2x\text{-}y}$, $C_{4z}\times\tilde{T}$, $C_{2z}\times\tilde{T}$, $C_{2x}\times\tilde{T}$, $C_{2y}\times\tilde{T}$, $C_{2xy}\times\tilde{T}$, $C_{2x\text{-}y}\times\tilde{T}$}, where $\tilde{T}$ represents the combined operation of time-reversal $T$ and different fractional lattice translation $\tau$, i.e., $T\tau$. As discussed in **METHODS**, $\tau$ can be omitted, allowing $T$ *vs* $T\tau$ to be treated equally. According to ***Rules I*** and ***IV***, ***S*(*k*)** must be parallel to the rotation axes and the $C_2\times T$-invariant planes in the corresponding symmetry-preserved *k*-invariant regions. The presence of $T$ establishes the relation $\boldsymbol{S}(\boldsymbol{k}) = -\boldsymbol{S}(-\boldsymbol{k})$. The four-fold rotation $C_{4z}$ ensures that ***S*(*k*)** for the two $C_{2x}$ and $C_{2y}$ axes (or the two $C_{2xy}$ and $C_{2x\text{-}y}$ axes) maintains the same magnitudes. Therefore, we can construct two types of symmetry-allowed spin frameworks to describe all possible 2D spin distributions with $422.1'$ MPG symmetry [**Figures 3H and 3K**]. Specifically, for the type-I spin framework [**Figure 3H**], we predict nontrivial STs in the 2D plane with parallel spin-momentum locking, provided the system has an isotropic constant-energy contour. **Figure 5C** shows the DFT-calculated radial ST for $ZnV_2O_4$. The presence of the $C_{2z}\times T$ operation guarantees a null $S_z$ spin component in the $k_z$=0 plane. Remarkably, compared to chiral Te[30,32], which has a significant out-of-plane $S_z$ component [Inset of **Figure 5C**], the radial ST in $ZnV_2O_4$ is much more ideal. The radial ST is reversed in $ZnV_2O_4$ with the opposite Néel-vector [**Figure S19**] rather than with opposite chirality[9,29,30].

For the latter case, we select the noncollinear AFM $Mn_3Ir$ to explore the physical origin of the peculiar shapes of STs. Notably, $Mn_3Ir$ is known for its large anomalous Hall effect, with such STs partially displayed in previous works[43] and believed to be related to topological superconductivity [44]. However, the origin of the peculiar shapes of STs in the [111] plane remains unclear. Here, we demonstrate that these STs can be understood using our ***Rules IV*** and ***V.*** The centrosymmetric $Mn_3Ir$ exhibits a triangular coplanar AFM ground-state (MSG $R\bar{3}m'$) [43,45], in which three adjacent magnetic Mn atoms follow an all-in-all-out pattern that breaks $PT$ operation, as shown in **Figure 5D. Figure 5E** displays their first Brillouin zone. The symmetry operations at the $\Gamma$-point include {$E$, $C_{3,111}$, $P$, $\bar{3}_z$, $C_{2,01\text{-}1}\times T$, $C_{2,1\text{-}10}\times T$, $C_{2,\text{-}101}\times T$, $m_{01\text{-}1}\times T$, $m_{1\text{-}10}\times T$, $m_{\text{-}101}\times T$ }. According to ***Rules IV*** and ***V***, ***S*(*k*)** is constrained to be parallel to the $C_2\times T$-invariant plane and perpendicular to the $m\times T$-invariant axis in the corresponding symmetry-preserved *k*-invariant regions. The presence of $P$ holds the relation $\boldsymbol{S}(\boldsymbol{k}) = \boldsymbol{S}(-\boldsymbol{k})$. Additionally, the three-fold rotation $C_{3,111}$ makes that the in-plane components of ***S*(*k*)** of three $C_2\times T$-invariant planes (three $m\times T$-invariant axes) hold the same magnitudes. Therefore, we can construct four types of symmetry-allowed frameworks to describe these in-plane spin distributions with $\bar{3}m'$ symmetry [**Figure S21**]. We note that Type-I (Type-III) can be transformed into Type-II (Type-IV) by flipping the Néel-vectors of the system. As illustrated in **Figure 5E,** we display DFT-calculated STs of the band near the Fermi-level, corresponding to the Type-I spin framework. If we neglect the out-of-plane components of ***S*(*k*)**, these tangential-radial-like patterns in the [111] plane strongly resemble traditional Dresselhaus STs [27], where ***S*(*k*)** is radial along the $k_x$ and $k_y$ axes and tangential to the diagonal directions [see inset of **Figure 3F**]. Accordingly, we may define a magnetic Dresselhaus effect to characterize the STs in $Mn_3Ir$ and other similar systems with $\bar{3}m'$ MPG symmetry.

### *Symmetry-constrained 3D persistent spin textures*

While ***Rule II*** indicates that 2D PSTs can appear in the mirror plane of crystals [31,41,46], the existence of symmetry-constrained 3D PSTs remains an unsolved mystery. In this study, we propose that two cases of symmetry-constrained 3D PSTs can exist for the first time by combining our rules with topological band theory, contributing to the design of spintronic devices with long spin decoherence time. Case-I is generated by a two-fold screw-rotation combined with a parallel fractional translation and the operation *T* (see left-panel, **Figure 6**). As illustrated in **Figure 6A**, we present a schematic diagram of case-I 3D PST: (i) According to ***Rule I***, strict parallel-spin-momentum locking occurs along the screw rotation axis. (ii) Based on TBT, a $s_{2,001} \times T$-enforced nodal plane is formed in the $k_z = \pi/c$ plane [47-51], where this two-fold degeneracy resembles Kramers' degeneracy, with nearly null spin polarization in the vicinity of high-symmetry points (such as A, L, and H points). (iii) When ***S*(*k*)** changes continuously as a function of perturbations with the same irreducible representation [33], ***S*(*k*)** deviating from the nodal plane can be viewed as a linear combination of ***S*(*k*)** along the screw-rotation[24]. Ultimately, ***S(k)*** deviating from the nodal plane will align uniformly along the screw-rotation axis, forming a 3D PST. In general, the 3D PST can be explained through the combination of our rules and the continuity assumption of ***S(k)***, independent of the establishment of a $k \cdot p$ Hamiltonian. This discussion can be also understood from the perspective of an effective *k*·p Hamiltonian, where the retained *k*-linear term $k_z\sigma_z$ dominates near the high-symmetry points. To validate our idea, we select nonmagnetic $PI_3$ as an example, with its first Brillouin zone shown in **Figure 6B**. There is a two-fold screw rotation $s_{2,001}$ along the *z*-axis, along with a $s_{2,001} \times T$-enforced nodal-plane at the $k_z = \pi$ plane. The calculated 3D STs of second-lowest conduction band surrounding the A point are shown in **Figure 6C**, exhibiting two features: i) 3D PSTs along the *z* direction exist within the full Bloch sphere where the *k*-linear term $k_z\sigma_z$ dominates; **Figure 6D** further shows a slice of the 3D PST in a 2D cut-plane. ii) The direction of ***S*(*k*)** flips from +*z* in $k_z > 0$ to -*z* in $k_z < 0$, due to *T* symmetry. Interestingly, the existence of 3D PSTs is independent of the specific band index and can be described by $k_z\sigma_z$ (**Figures S25-S26**), which are also observed around the L and H points (**Figure S24**). The 3D PST can be maintained within a large radius of 0.1 $\text{Å}^{-1}$ (**Figure S27**). Due to a similar origin, Case-I 3D PSTs are also found in the nonmagnetic $Ag_2Se$ (**Figure S32**).

Case-II 3D PSTs are generated by glide-mirror that feature at least a parallel fractional translation combined with *T* (see right panel, **Figure 6**). As shown in **Figure 6E**, the schematic diagram of case-II 3D PSTs shows: (i) According to ***Rule II***, strict perpendicular-spin-momentum locking exists within the glide mirror plane. (ii) Based on TBT, two $\tilde{m}_{010} \times T$-enforced nodal lines are formed along the *y* direction in the $k_x = \pi/a$ plane [49-52]. The fractional translation determines the positions of these nodal lines, and this two-fold degeneracy is similar to Kramers' degeneracy with nearly null spin polarization near the high-symmetry points (such as U, R, X, and S points). (iii) Similarly, when ***S*(*k*)** changes continuously as a function of perturbations[33], ***S*(*k*)** deviating from the nodal line can be viewed as a linear combination of ***S(k)*** in the glide-mirror plane[24]. Consequently, ***S(k)*** within a certain range will be uniformly aligned perpendicular to the glide-mirror plane, forming a 3D PST. Additionally, the above discussion can be also understood from the perspective of effective *k*·p Hamiltonian, in which the only retained *k*-linear term $k_x\sigma_y$ is dominated. Again, to confirm our design principle, we select nonmagnetic PbS as an example, whose first Brillouin zone is shown in **Figure 6F**. There is a glide-mirror $\tilde{m}_{010}$ perpendicular to $y$ axis and two nodal-lines along U-R and X-S in the Brillouin zone. The calculated 3D STs of lowest conduction band surrounding R point are shown in **Figure 6G**, which also exhibit two features: i) 3D PSTs along $y$ direction exist in the full Bloch sphere where $k$-linear term $k_x\sigma_y$ is dominated; **Figure 6H** further shows the slice of 3D PST in 2D cut-plane. ii) The direction of ***S*(*k*)** flips from +$y$ in $k_x < 0$ to -$y$ in $k_x > 0$. The existence of 3D PSTs is independent of the specific energy band and can be described by $k_x\sigma_y$ (**Figures S29-S30**), which are also found around X, S and U points (**Figure S28**) and almost maintained within a large radius of 0.1 $\text{Å}^{-1}$ (**Figure S31**). Besides of PbS, Case-II 3D PSTs are also observed in nonmagnetic $Ba_2Ti(GeO_4)_2$ (**Figure S33**). It is noted that in a system with more than one non-symmorphic symmetry, the additional presence of $s_2$ or $\tilde{m}$ will affect the band-crossing at the specific boundary-points of Brillouin zone [53], in which 3D PSTs could be broken (**Figures S32-S33**).

Furthermore, we acknowledge that Tao *et. al*[24] initially demonstrated the glide-mirror protected 2D PSTs around the X and Y points of orthorhombic non-symmorphic crystals with the $C_{2v}$ point group. Our study reveals that two cases of 3D PSTs can exist across a broader range of materials (including those with $C_6$, $D_2$, and $C_{4v}$ point groups) and a wider range of Brillouin zone (such as U, R, and S points), achieved by combining our symmetry-constrained rules with known TBT results and the continuity assumption of ***S(k)***.

## DISCUSSION

We note that while ***Rules I-II*** may be partially indicated in the literature [32,41], ***Rules III-V*** are proposed for the first time in this study. Importantly, we emphasize that ***Rules III-V*** are not simply extensions of ***Rules I-II*** that incorporate time-reversal symmetries. Unlike ***Rules I-II***, ***Rules III-V*** operate in completely different momentum subspace and correspond to entirely different symmetry operations. For example, although $m$ operation can be derived from the rotation $C_2$ and space-inversion $P$ ($m=P\times C_2$), the system holding $m$ operation is distinctly different from the system that holds both $C_2$ and $P$ operations. The former corresponds to the non-centrosymmetric $C_s$ point group, while the latter corresponds to the centrosymmetric $C_{2h}$ point group. Overall, ***Rules I-V*** result from a complete symmetry analysis in which we enumerated all symmetry operations in the period solids with SOC.

We also note that the $k{\cdot}p$ method is quite general and universal, allowing for the construction of suitable effective $k{\cdot}p$ Hamiltonians to explain various physical effects, including spin-polarization effects, both analytically and numerically, according to specific situations[32-35]. However, the challenge of the $k{\cdot}p$ method lies in establishing appropriate $k{\cdot}p$ Hamiltonians that can accurately describe a particular material. Even at the same high-symmetry point of a material, electronic states with different band indices can take different forms of $k{\cdot}p$ Hamiltonians. This process typically demands considerable expertise and experience, as it involves various factors, including atomic orbitals, the selection of appropriate basis vectors, band representations, and Wyckoff positions. While principles for establishing the $k{\cdot}p$ Hamiltonian[33,35] can be provided, it remains challenging to enumerate all the different forms of effective Hamiltonians that explain various physical effects in a straightforward manner. In contrast, our rules offer a complementary and independent perspective for understanding the diverse and intricate shapes of STs in *PT*-broken crystals, including 2D materials. One can establish preliminary spin frameworks for a material simply and efficiently by relying solely on its (magnetic) point group symmetry. These preliminary frameworks are typically sufficient for understanding the spin-polarization properties when combined with the continuity assumption of ***S*(*k*)**.

For the direct application of our rules, we provide a systematic understanding of all ZSPLs in crystals, while case-II of pinned ZSPLs has been discussed in the literature[40]. The classification of ZSPLs has not been performed before and is proposed for the first time based on ***Rules I-III***. Next, we predict vertex-like and windmill-like STs in crystals with $D_n$ and $C_{nv}$ point groups, respectively, which motivates a deeper understanding of the spin-polarization effects from a purely symmetry-based perspective. We also demonstrate that radial STs with parallel spin-momentum locking can emerge in conventional AFM material $ZnV_2O_4$, where the band splitting originates entirely from the inclusion of SOC. Finally, we illustrate that the symmetry-constrained 3D PSTs can appear in two families of non-symmorphic crystals for the first time, remaining robust against symmetry-allowed impurities [54]. Compared to symmetry-constraint 2D PSTs [24,31], we expect that 3D PSTs can provide a more stable electronic state to produce a persistent spin helix. We also note that our rules and discussions can be precisely applied to understanding diverse orbital textures in crystals[55-57], providing fundamental principles for variable spintronic and orbitronic applications.

## METHODS

### *First-principles calculations*

All first-principles density-functional theory (DFT) calculations were performed by the VASP software [58,59] with the Perdew-Burke-Ernzerhof (PBE) exchange-correlation functional [60]. The structures were fully relaxed until the residual force acting on each ion was less than 0.01 eV/Å and the total energy was converged up to $10^{-6}$ eV/cell. The plane-wave cutoff energy was set to 1.5×ENMAX for different material systems and the Monkhorst-Pack *k* grid was set to 2π × 0.02 $Å^{-1}$ for Brillouin zone sampling. The spin-orbit coupling (SOC) was included in the first-principles calculations. The primitive-cell structures of nonmagnetic materials and magnetic materials are accessible through the Topological Materials Database [49-51] and the MAGNDATA database [61], respectively.

### *Derivation of basic rules describing the spin-polarization effect in PT-broken systems*

Firstly, we explore the symmetry operations in three-dimensional solids with spin-orbit coupling (SOC), ultimately simplifying them to four basic operations. It is believed that the symmetries of three-dimensional periodic solids with SOC can be described by a complete crystallographic group theory, including 32 point groups (PGs), 230 space groups (SGs), 122 magnetic PGs (MPGs), 1651 magnetic SGs (MSGs), and their double groups. In **Tables S1 and S2** of **Note.S1** of Supplementary Information, we enumerate the symmetry operations for 32 PGs and 122 MPGs respectively, through Bilbao Crystallographic Server [62]. According to **Tables S1 and S2**, we find that 32 PGs consist of the spatial symmetries, including space inversion (*P*), rotations ($C_n$, *n*=2, 3, 4 and 6), mirror (*m*) and roto-inversions ($\bar{n}$, *n*=3, 4 and 6), while 122 MPGs include not only the spatial symmetries mentioned above but also the anti-unitary symmetry, i.e., time reversal (*T*), and their combinations with *T.* Based on 32 PGs and 122 MPGs, considering both integer and fractional crystal translations ($\tau$) yields 230 SGs and 1651 MSGs. Additionally, we noted that $m$ and $\bar{n}$ can be derived from combining twofold rotation with space inversion, and $n$-fold rotation with space inversion, respectively, i.e., $m = P \times C_2$ and $\bar{n} = P \times C_n$ ($n$=3, 4 and 6). Therefore, we can classify all the symmetry operations into four basic symmetry operations, *i.e.*, space inversion (*P*), time reversal (*T*), integer and fractional lattice translations ($\tau$) and rotations ($C_n$).

Next, we explore the role of the symmetry operations on the spin polarization vector $\boldsymbol{S}(\boldsymbol{k})$ in spin-orbit couped systems. First of all, the presence of SOC breaks the spinor operation *U* of SU(2) group, in which *U* satisfies $U\boldsymbol{S}(\boldsymbol{k}) = -\boldsymbol{S}(\boldsymbol{k})$. Since *P* reverses the coordinate $\boldsymbol{r}$ ($P\boldsymbol{r} = -\boldsymbol{r}$), *T* reverses the time $\boldsymbol{t}$ ($T\boldsymbol{t} = -\boldsymbol{t}$), and $\tau$ shift $\boldsymbol{r}$ ($\tau\boldsymbol{r} = \boldsymbol{r} + \boldsymbol{a}/n$, *a*=lattice vector, *n*=positive integer), *P*, *T* and $\tau$ operations act on the spin polarization vector $\boldsymbol{S}(\boldsymbol{k})$ as:

$$P\boldsymbol{S}(\boldsymbol{k}) = \boldsymbol{S}(-\boldsymbol{k}) \qquad \text{(Equation 1)}$$

$$T\boldsymbol{S}(\boldsymbol{k}) = -\boldsymbol{S}(-\boldsymbol{k}) \qquad \text{(Equation 2)}$$

$$\tau\boldsymbol{S}(\boldsymbol{k}) = \boldsymbol{S}(\boldsymbol{k}) \qquad \text{(Equation 3)}$$

Therefore, the combination of *PT or PT*$\tau$ preserves double spin degeneracy for arbitrary $\boldsymbol{k}$ in the spin-orbit coupled system, according to **Equations 1-3**. Since $\tau\boldsymbol{S}(\boldsymbol{k}) = \boldsymbol{S}(\boldsymbol{k})$, we would omit the symbol $\tau$ for convenience in the following sections. In the other words, we would treat *PT vs PT*$\tau$, and *T vs T*$\tau$ equally. It is noted that $T\tau$ operation corresponds to Type-IV MSGs, in which time-reversal operation is broken.

For *PT*-broken systems with SOC, the breaking of *PT* symmetry results in nonzero spin polarization, and the presence of SOC locks spin rotation and spatial rotation completely, in which a spatial rotation $C_n$ is accompanied by an identical spin rotation[63]. As a result, $C_{nz}$ operation acts on the spin polarization vector $\boldsymbol{S}(\boldsymbol{k})$ as:

$$C_{nz}\boldsymbol{S}(\boldsymbol{k}) = (C_{nz}S_x, C_{nz}S_y, S_z)(C_{nz}k_x, C_{nz}k_y, k_z) \qquad \text{(Equation 4)}$$

Additionally, since $m$=$P\times C_2$ and $\bar{n}$= $P\times C_n$ ($n$=3, 4 and 6), $m_z$ and $\bar{n}_z$ operations act on the spin polarization vector $\boldsymbol{S}(\boldsymbol{k})$ as:

$$m_z\boldsymbol{S}(\boldsymbol{k}) = PC_{2z}\boldsymbol{S}(\boldsymbol{k}) = (-S_x, -S_y, S_z)(k_x, k_y, -k_z) \qquad \text{(Equation 5)}$$

$$\bar{n}_z\boldsymbol{S}(\boldsymbol{k}) = PC_{nz}\boldsymbol{S}(\boldsymbol{k}) = (C_{nz}S_x, C_{nz}S_y, S_z)(-C_{nz}k_x, -C_{nz}k_y, -k_z) \qquad \text{(Equation 6)}$$

Combining the spatial operations with anti-unitary operation *T*, the corresponding transformations follow:

$$C_{nz} \times T\boldsymbol{S}(\boldsymbol{k}) = (-C_{nz}S_x, -C_{nz}S_y, -S_z)(-C_{nz}k_x, -C_{nz}k_y, -k_z) \qquad \text{(Equation 7)}$$

$$m_z \times T\boldsymbol{S}(\boldsymbol{k}) = (S_x, S_y, -S_z)(-k_x, -k_y, k_z) \qquad \text{(Equation 8)}$$

$$\bar{n}_z \times T\boldsymbol{S}(\boldsymbol{k}) = (-C_{nz}S_x, -C_{nz}S_y, -S_z)(C_{nz}k_x, C_{nz}k_y, k_z) \qquad \text{(Equation 9)}$$

The observable spin-polarization vector ***S*(*k*)** is severely constrained by the symmetry *O* on *M[O]* ($M[O] = \{\boldsymbol{k}|O\boldsymbol{k} = \boldsymbol{k}\}$) [37]. Besides the isolated time-reversal invariant momentum (TRIM) points (They satisfy $\boldsymbol{k} = T\boldsymbol{k} = P\boldsymbol{k} = -\boldsymbol{k}$), according to **Equations 4-9**, we can only obtain five types of *M[O]*, *i.e.*, rotation axis (such as $C_{nz}$, *i.e.*, $k_z$ axis), mirror plane (such as $m_z$, *i.e.*, $k_z$=0/$\pi$ plane), $\bar{n} \times T$-invariant axis (such as $\bar{n}_z \times T$, *i.e.*, $k_z$ axis), $C_2 \times T$-invariant plane (such as $C_{2z} \times T$, *i.e.*, $k_z$=0/$\pi$ plane) and $m \times T$-invariant axis (such as $m_z \times T$, *i.e.*, $k_z$ axis). Therefore, the spin polarization vector ***S*(*k*)** would satisfy such five basic rules.

We note that **Equations 1-9** can be derived by considering the wavefunction origin of ***S*(*k*)**. The constraints on ***S*(*k*)** can be explained by examining the evolution of spinful wavefunctions under the symmetry operations. As an example, we use the two-fold rotation operation $C_{2z}$ to demonstrate it. First, ***S*(*k*)** of a specific band can be calculated by

$$\langle S_i^n(\boldsymbol{k})\rangle = 1/2\,\langle u_n(k)|\hat{\sigma}_i|u_n(k)\rangle,\ i = x, y, z \qquad \text{(Equation 10)}$$

Here, $\hat{\sigma}_i$ and $u_n(k)$ represents the Pauli matrix and the spinful wavefunction of *n*th nondegenerate band, respectively. Then, we have the relation under the symmetry,

$$\begin{aligned} C_{2z}\langle S_x^n(\boldsymbol{k})\rangle &= 1/2\langle u_n(k)|C_{2z}^{-1}\, C_{2z}\hat{\sigma}_x C_{2z}^{-1}\, C_{2z}|u_n(k)\rangle \\ &= 1/2\langle u_n(-k_x, -k_x, k_z)|-\hat{\sigma}_x|u_n(-k_x, -k_x, k_z)\rangle \\ &= -\langle S_x^n(-k_x, -k_x, k_z)\rangle \qquad \text{(Equation 11)} \end{aligned}$$

Similarly, we have the relations for the other components, $C_{2z}\langle S_y^n(k)\rangle = -\langle S_y^n(-k_x, -k_x, k_z)\rangle$ and $C_{2z}\langle S_z^n(k)\rangle = \langle S_z^n(-k_x, -k_x, k_z)\rangle$. Then, by omitting the *n* and bra-ket symbols for simplicity, we obtain the following relation,

$$C_{2z}\boldsymbol{S}(\boldsymbol{k}) = (-S_x, -S_y, S_z)(-k_x, -k_y, k_z) \qquad \text{(Equation 12)}$$

Here, $S_i$ represents the expectation value of the component of ***S*(*k*)**. we note that **Equation 12** corresponds to **Equation 4**. Additionally, one can derive **Equations 1-9** of the manuscript from the above similar procedure.

## RESOURCE AVAILABILITY

### *Lead contact*

Further information and requests for resources and reagents should be directed to and will be fulfilled by the lead contact, Bing Huang (bing.huang@csrc.ac.cn).

### *Materials availability*

This study did not generate new materials.

### *Data and code availability*

Any additional information required to reanalyze the data reported in this paper is available from the lead contact upon request.

## ACKNOWLEDGMENTS

This work is supported by NSFC (Grants Nos. 12088101, 12474217 and 12004030), the National Key Research and Development of China (Grant No. 2022YFA1402401), NSAF (Grant No. U2230402), Beijing Natural Science Foundation (Grant No. 1242023) and the Fundamental Research Funds for the Central Universities. Calculations were done in Tianhe-JK cluster at CSRC.

## AUTHOR CONTRIBUTIONS

W.T. and B.H. conceived the project. W.T. carried out the DFT calculations and formula derivation. W.T., JF.W., Y.L. and B.H. performed the symmetry analysis. W.T., JF.W., and B.H. wrote the paper. All authors discussed the results.

## DECLARATION OF INTERESTS

The authors declare no competing interests.

## SUPPLEMENTAL INFORMATION

This file contains detailed information on all symmetry operations of the 32 point groups and 122 magnetic point groups, as well as data on the crystal structures, Brillouin zones, band dispersions, other spin textures, and the detailed analysis of 3D persistent spin textures.

**Document S1. Notes S1-S5, Figures S1–S33, and Tables S1-S3**

## FIGURE TITLES AND LEGENDS

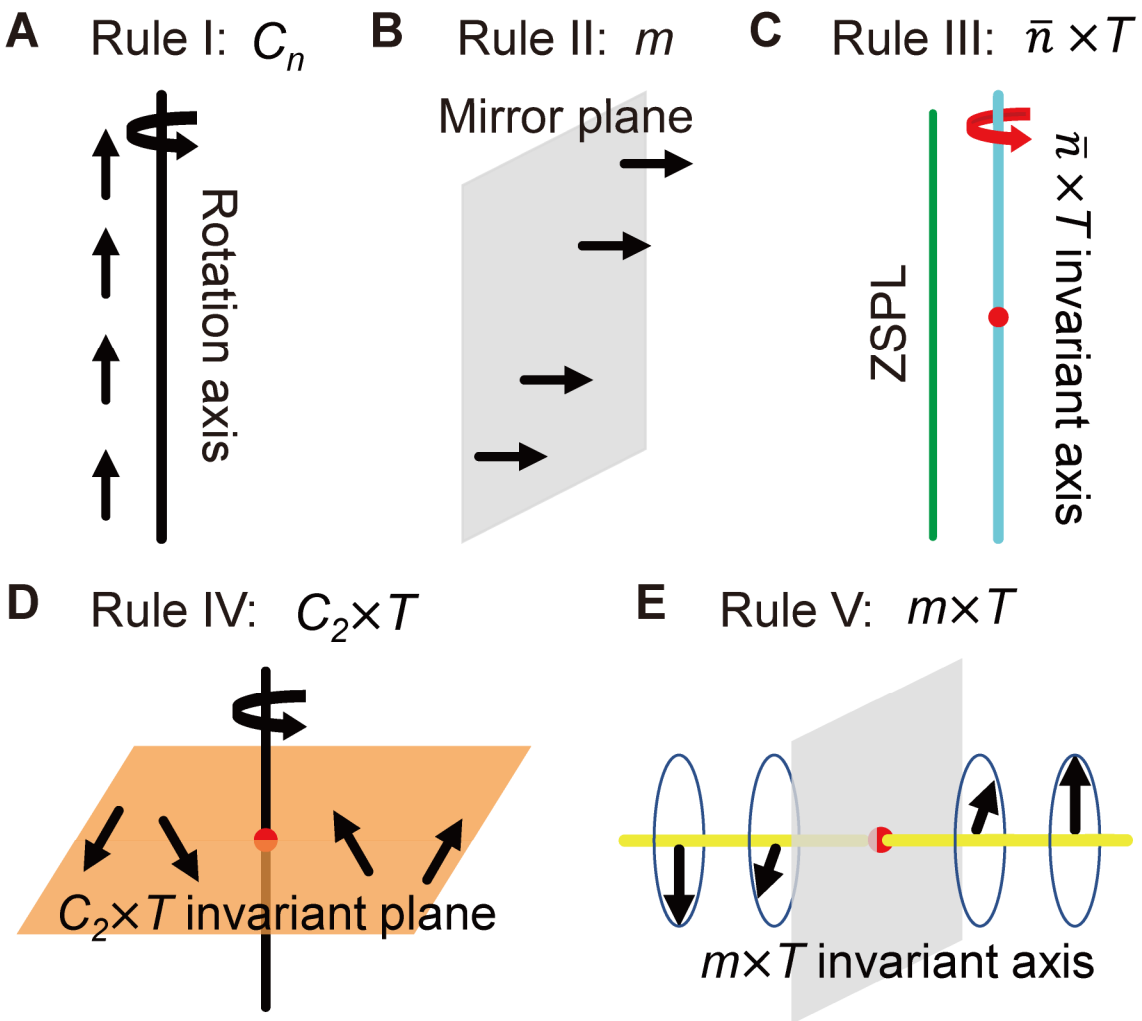

**Figure 1. Symmetry-constrained rules of spin-polarization**

(A) *Rule I*: along rotation axis ($C_n$), only parallel spin-polarization component can exist.

(B) *Rule II*: in mirror plane ($m$), only out-of-plane spin-polarization component can exist.

(C) *Rule III*: along $\bar{n}\times T$-invariant axis ($\bar{n}$ represents roto-inversion), a zero spin polarization line (ZSPL) is obtained (denoted by green-line).

(D) *Rule IV*: in $C_2\times T$-invariant plane, only in-plane spin-polarization component can exist.

(E) *Rule V*: along $m\times T$-invariant axis, only perpendicular spin-polarization component can exist.

Arrows indicate the directions of spin-polarization vectors. Grey and orange planes represent mirror and $C_2\times T$-invariant planes, respectively. Black, cyan and yellow lines represent rotation, $\bar{n}\times T$-invariant and $m\times T$-invariant axes, respectively.

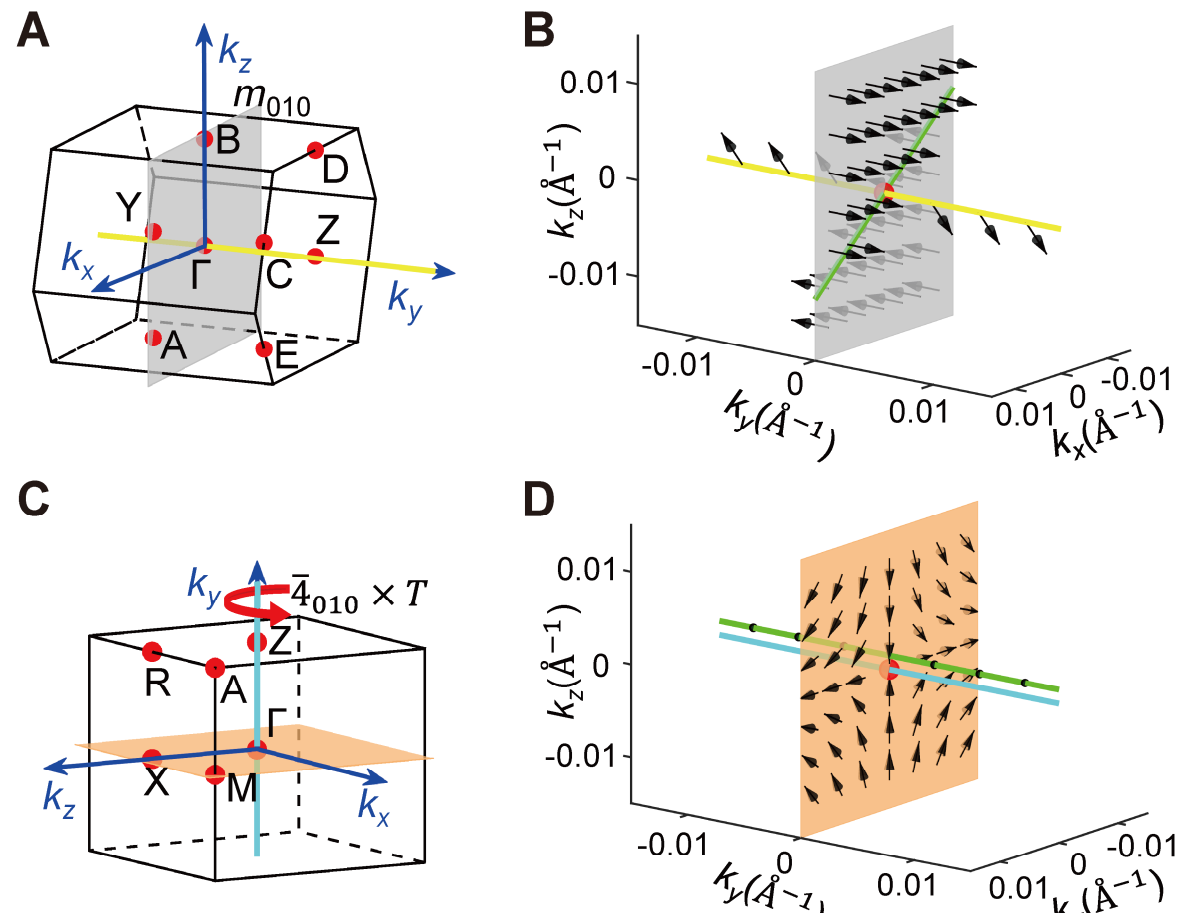

**Figure 2. Complete classification of zero spin polarization lines**

(A and B) First Brillouin zone and spin texture with unpinned ZSPL (denoted by green-line) in $ReN_2$ (space group *P*m).

(C and D) First Brillouin zone and spin texture with pinned ZSPL in $GeSe_2$ (space group $P\bar{4}$).

Red-dots represent time-reversal-invariant-momenta. Here, spin textures are plotted for the band near Fermi-level centered around $\Gamma$-point. Grey and orange planes represent mirror and $C_2\times T$-invariant planes, respectively. Cyan and yellow lines represent $\bar{n}\times T$-invariant and $m\times T$-invariant axes, respectively.

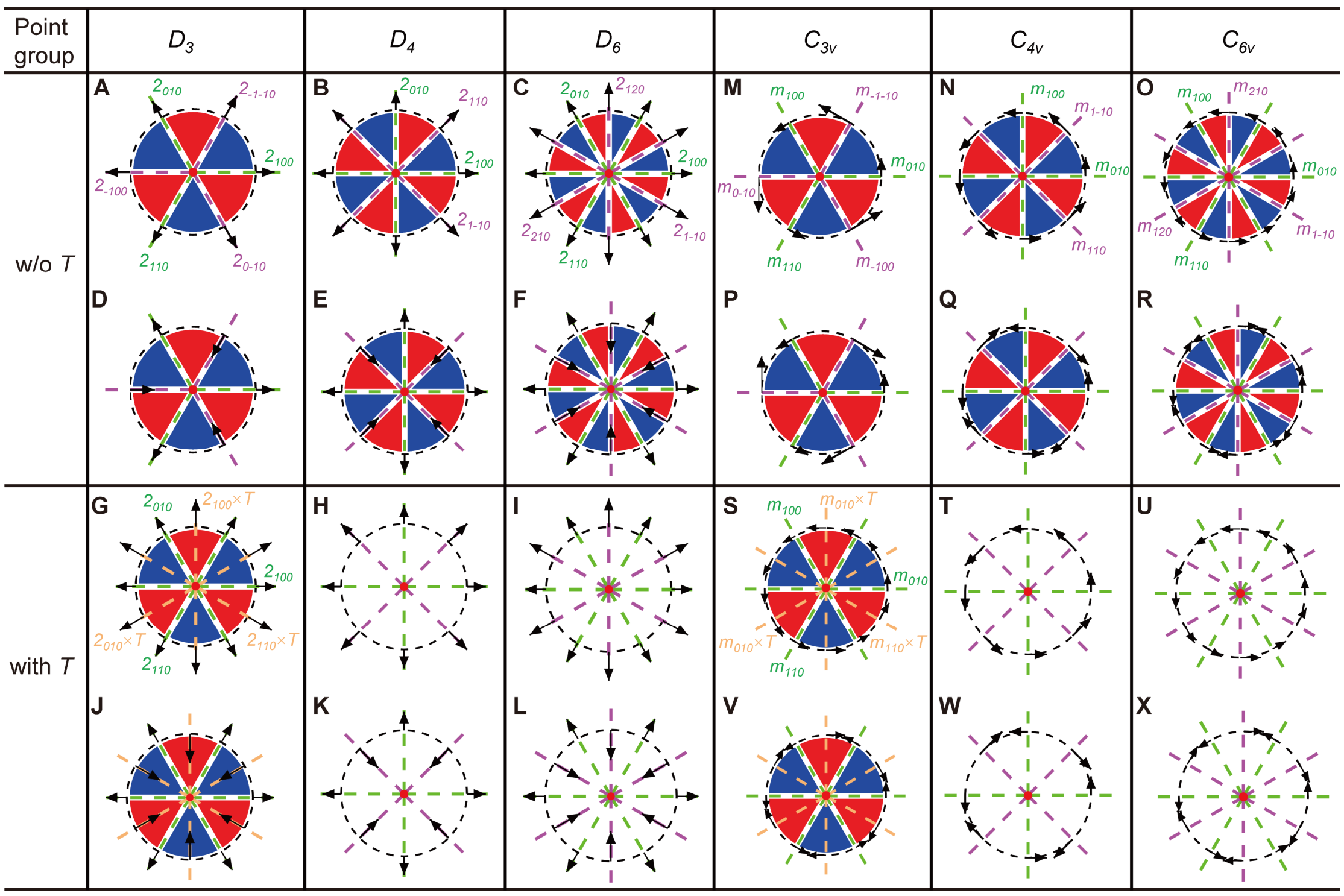

**Figure 3. Two distinct types of spin vector field with $D_n$ and $C_{nv}$ point groups ($n$=3, 4 and 6)**

(A-C and G-I) For type-I spin frameworks with $D_n$ point group, a radial-like spin texture can exist.

(D-F and J-L) For type-II spin frameworks with $D_n$ point group, a vertex-like spin texture is predicted.

(M-O and S-U) For type-I spin frameworks with $C_{nv}$ point group, a Rashba-like spin texture can exist.

(P-R and V-X) For type-II spin frameworks with $C_{nv}$ point group, a windmill-like spin texture is predicted.

The arrows represent the spin-polarization vectors ***S*(*k*)** of the rotation axes/mirror planes. The green- and purple- dashed-lines represent the two types of in-plane two-fold rotation axes (vertical mirror planes), where ***S*(*k*)** can exhibit different magnitudes and orientations. The (red/blue) color of the planes represents the nonzero out-of-plane (up/down) spin components. Additionally, the orange dashed-lines represent the vertical $C_2 \times T$ or in-plane $m \times T$ symmetries [(G), (J), (S) and (V)]. It is noted that the in-plane rotation axes ($m \times T$ axes) also lied in the vertical $C_2 \times T$ planes (mirror planes) for $D_4$/$D_6$ ($C_{4v}$/$C_{6v}$) point groups with $T$ symmetry [(H)-(I), (K)-(L), (T)-(U) and (W)-(X)].

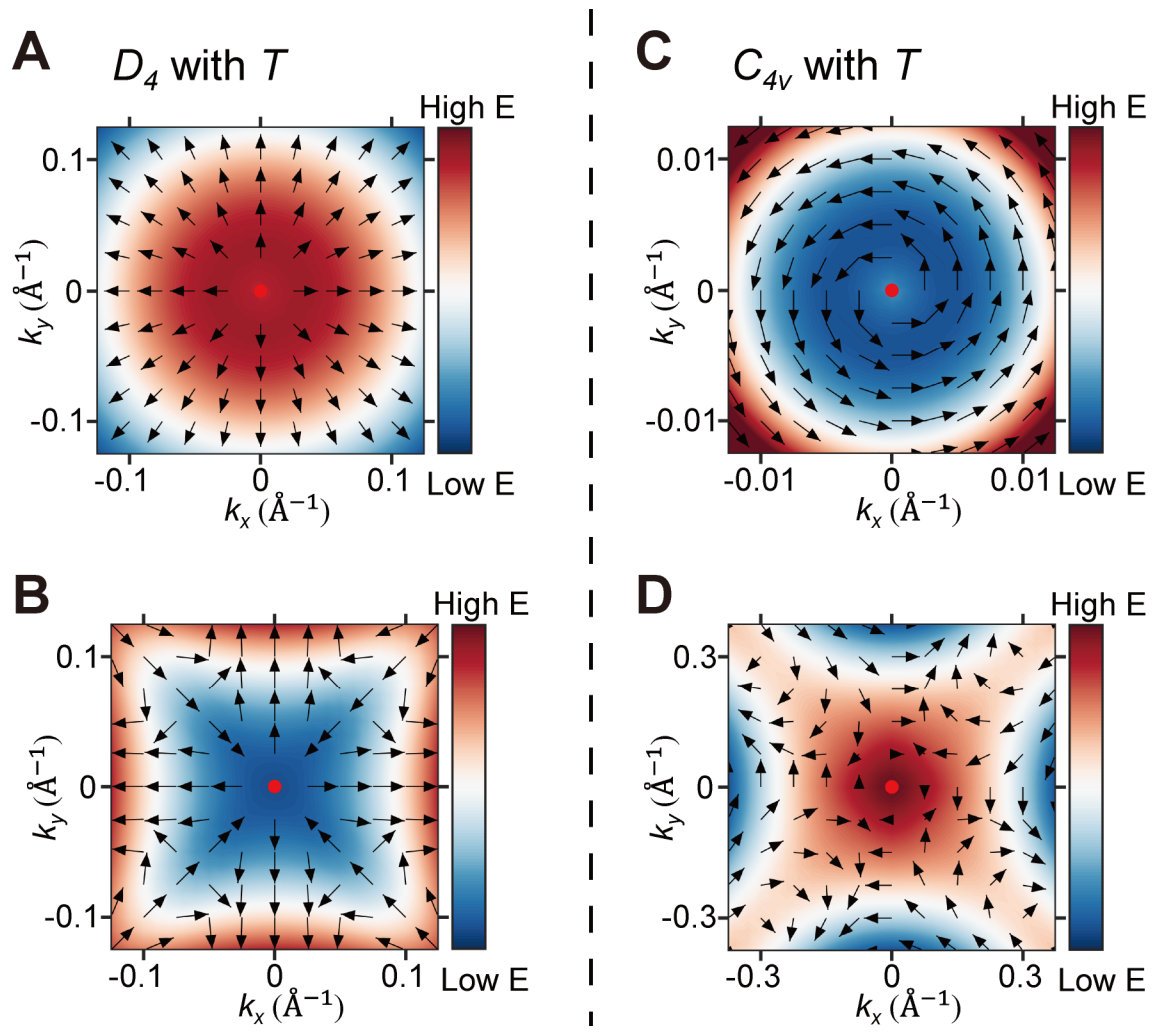

**Figure 4. Spin vector fields in nonmagnetic crystals with $D_4$/$C_{4v}$ point group**

(A and B) spin textures of $TeO_2$ (space group $P4_32_12$) for the two typical bands with isotropic and anisotropic constant-energy contours, respectively.

(C and D) Same as (A)-(B) but for $AlN_2OTa$ (space group $P4mm$).

The red-dot represents the Γ point (see the crystal structures and energy dispersions in **Figures S16-S17**). The (red/blue) color of the planes represents the electronic states with (high/low) energy.

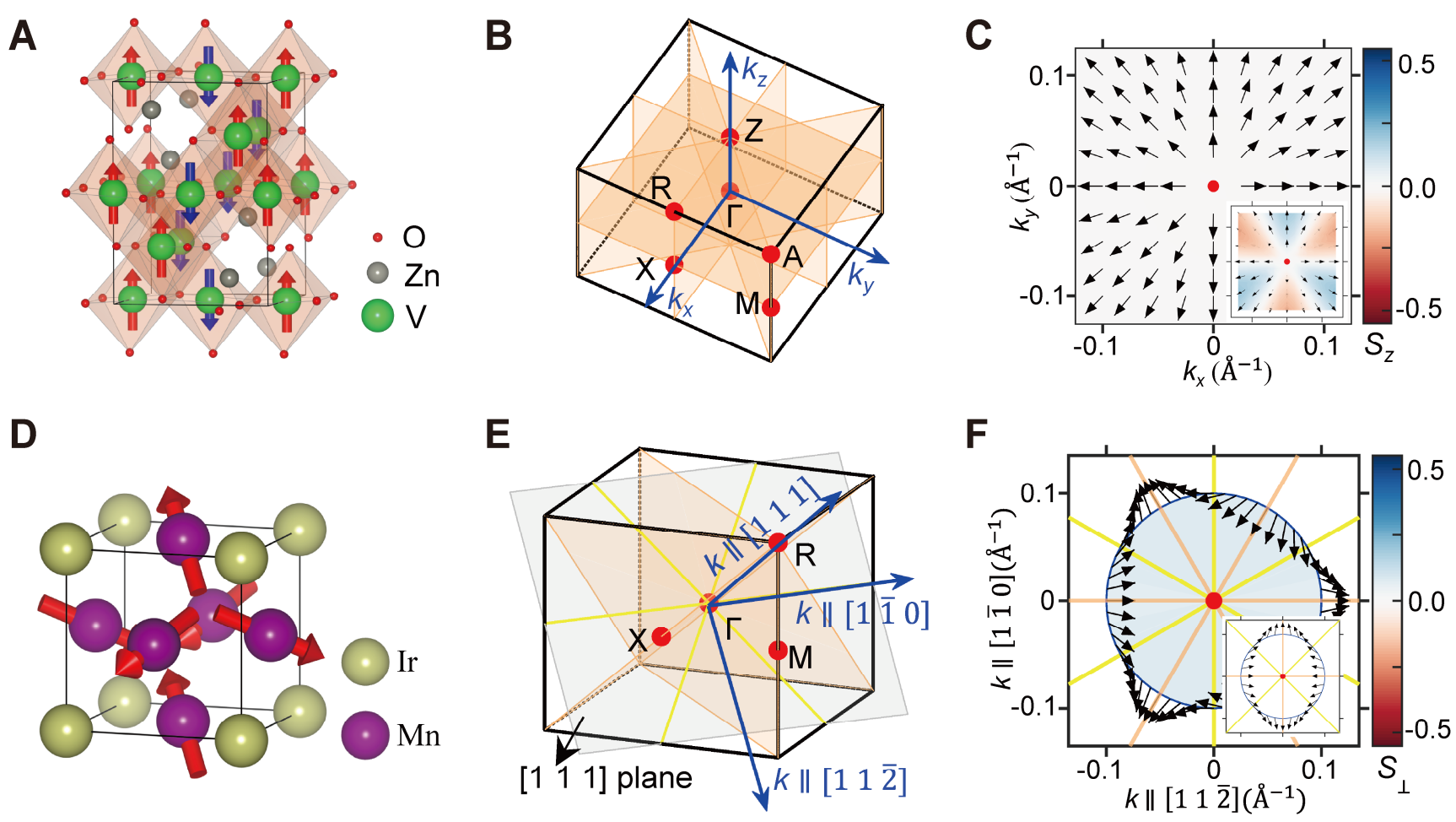

**Figure 5. Radial Spin texture and magnetic Dresselhaus effect in centrosymmetric crystals**

(A-B and D-E) Magnetic unit cell and first Brillouin zone for AFM-phase $ZnV_2O_4$ (space group $I4_1/amd$, magnetic space group $P_I4_32_12$) and $Mn_3Ir$ (space group $Pm\bar{3}m$, magnetic space group $R\bar{3}m'$). In (B) and (E), the orange-planes and yellow-lines represent $C_2 \times T$-invariant planes and $m \times T$-invariant lines, respectively. Arrows of (A) and (D) represent the orientation of magnetic moments.

(C and F) DFT-calculated spin textures around Γ-point are plotted for [001] plane in $ZnV_2O_4$ and [111] plane in $Mn_3Ir$. For comparison, insets of (C) and (F) show the spin textures in [001] plane around A-point of Te and Γ-point of $Ag_2HgI_4$, respectively (see details in **Figures S22-S23**). In (F), the orange-line represent the intersecting line of three $C_2 \times T$-invariant planes and [111] plane.

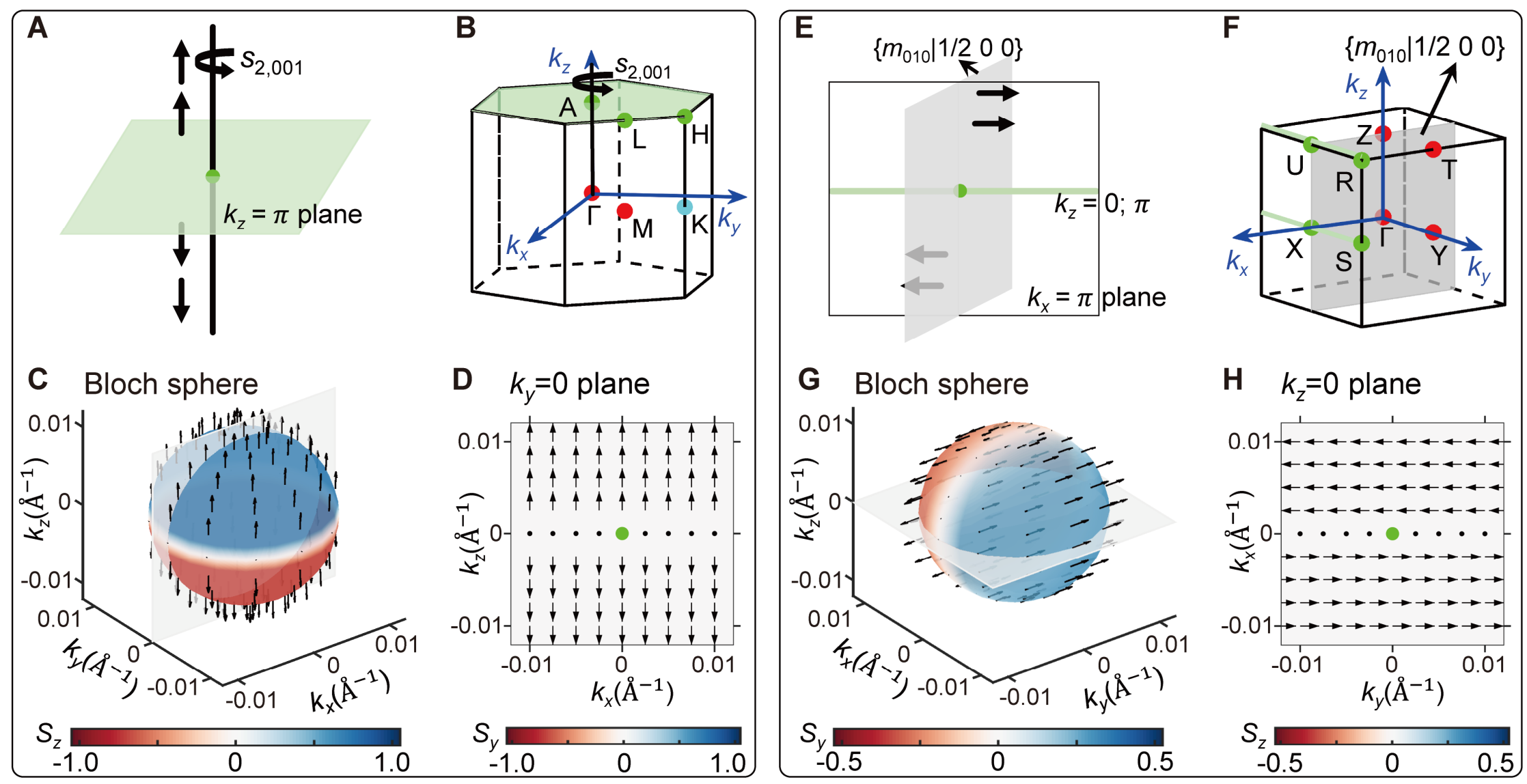

**Figure 6. Two cases of symmetry-constrained 3D persistent spin textures**

(A) Principle of Case-I 3D persistent spin texture (PST) produced by two-fold screw-rotation in combination with *T*.

(B) First Brillouin zone of material example $PI_3$ (space group $P6_3$). A twofold screw-rotation $s_{2,001} = \{C_{2,001}|\ 0\ 0\ 1/2\}$ is along *z* axis. Green-plane represents a nodal-plane.

(C) DFT-calculated 3D PST surrounding A point, where spin-polarization vectors are along *z* direction.

(D) PST from an arbitrary 2D cut-plane marked in (C).

(E) Principle of Case-II 3D PST produced by glide-mirror in combination with *T*.

(F) First Brillouin zone of material example PbS (space group *P*ma2). A glide mirror $\tilde{m}_{010} = \{m_{010}|\ 1/2\ 0\ 0\}$ perpendicular to *y* axis exists. Green-lines represent the nodal-lines.

(G) DFT-calculated 3D PST surrounding R point, where spin-polarization vectors are along *y* direction.

(H) PST from an arbitrary 2D cut-plane marked in (G).

Green-dots of panels (B) and (F) represent high-symmetry points generating 3D PSTs.

## REFERENCES

1. Zhang, X., Liu, Q., Luo, J.-W., Freeman, A.J., and Zunger, A. (2014). Hidden spin polarization in inversion-symmetric bulk crystals. Nat. Phys. *10*, 387-393. https://doi.org/10.1038/nphys2933.
2. Yuan, L., Liu, Q., Zhang, X., Luo, J.-W., Li, S.-S., and Zunger, A. (2019). Uncovering and tailoring hidden Rashba spin–orbit splitting in centrosymmetric crystals. Nat. Comm. *10*, 906. https://doi.org/10.1038/s41467-019-08836-4.
3. Yuan, L.-D., Zhang, X., Acosta, C.M., and Zunger, A. (2023). Uncovering spin-orbit coupling-independent hidden spin polarization of energy bands in antiferromagnets. Nat. Comm. *14*, 5301. https://doi.org/10.1038/s41467-023-40877-8.
4. Guan, S., Luo, J.-W., Li, S.-S., and Zunger, A. (2023). Hidden Zeeman-type spin polarization in bulk crystals. Phys. Rev. B *107*, L081201. https://doi.org/10.1103/PhysRevB.107.L081201.
5. Manchon, A., Koo, H.C., Nitta, J., Frolov, S.M., and Duine, R.A. (2015). New perspectives for Rashba spin–orbit coupling. Nat. Mater. *14*, 871-882. https://doi.org/10.1038/nmat4360.
6. Edelstein, V.M. (1990). Spin polarization of conduction electrons induced by electric current in two-dimensional asymmetric electron systems. Solid State Commun. *73*, 233-235. https://doi.org/10.1016/0038-1098(90)90963-C.
7. Sánchez, J.C.R., Vila, L., Desfonds, G., Gambarelli, S., Attané, J.P., De Teresa, J.M., Magén, C., and Fert, A. (2013). Spin-to-charge conversion using Rashba coupling at the interface between non-magnetic materials. Nat. Comm. *4*, 2944. https://doi.org/10.1038/ncomms3944.
8. Shen, K., Vignale, G., and Raimondi, R. (2014). Microscopic Theory of the Inverse Edelstein Effect. Phys. Rev. Lett. *112*, 096601. https://doi.org/10.1103/PhysRevLett.112.096601.
9. Calavalle, F., Suárez-Rodríguez, M., Martín-García, B., Johansson, A., Vaz, D.C., Yang, H., Maznichenko, I.V., Ostanin, S., Mateo-Alonso, A., Chuvilin, A., et al. (2022). Gate-tuneable and chirality-dependent charge-to-spin conversion in tellurium nanowires. Nat. Mater. *21*, 526-532. https://doi.org/10.1038/s41563-022-01211-7.
10. Cai, L., Yu, C., Zhao, W., Li, Y., Feng, H., Zhou, H.-A., Wang, L., Zhang, X., Zhang, Y., Shi, Y., et al. (2022). The Giant Spin-to-Charge Conversion of the Layered Rashba Material BiTeI. Nano Lett. *22*, 7441-7448. https://doi.org/10.1021/acs.nanolett.2c02354.
11. Bernevig, B.A., and Vafek, O. (2005). Piezo-magnetoelectric effects in p-doped semiconductors. Phys. Rev. B *72*, 033203. https://doi.org/10.1103/PhysRevB.72.033203.
12. Chernyshov, A., Overby, M., Liu, X., Furdyna, J.K., Lyanda-Geller, Y., and Rokhinson, L.P. (2009). Evidence for reversible control of magnetization in a ferromagnetic material by means of spin–orbit magnetic field. Nat. Phys. *5*, 656-659. https://doi.org/10.1038/nphys1362.

13. Manchon, A., Železný, J., Miron, I.M., Jungwirth, T., Sinova, J., Thiaville, A., Garello, K., and Gambardella, P. (2019). Current-induced spin-orbit torques in ferromagnetic and antiferromagnetic systems. Rev. Mod. Phys. *91*, 035004. https://doi.org/10.1103/RevModPhys.91.035004.
14. Grimaldi, E., Krizakova, V., Sala, G., Yasin, F., Couet, S., Sankar Kar, G., Garello, K., and Gambardella, P. (2020). Single-shot dynamics of spin–orbit torque and spin transfer torque switching in three-terminal magnetic tunnel junctions. Nat. Nanotechnol. *15*, 111-117. https://doi.org/10.1038/s41565-019-0607-7.
15. Ganichev, S.D., Ivchenko, E.L., Bel'kov, V.V., Tarasenko, S.A., Sollinger, M., Weiss, D., Wegscheider, W., and Prettl, W. (2002). Spin-galvanic effect. Nature *417*, 153-156. https://doi.org/10.1038/417153a.
16. Benítez, L.A., Savero Torres, W., Sierra, J.F., Timmermans, M., Garcia, J.H., Roche, S., Costache, M.V., and Valenzuela, S.O. (2020). Tunable room-temperature spin galvanic and spin Hall effects in van der Waals heterostructures. Nat. Mater. *19*, 170-175. https://doi.org/10.1038/s41563-019-0575-1.
17. Khokhriakov, D., Hoque, A.M., Karpiak, B., and Dash, S.P. (2020). Gate-tunable spin-galvanic effect in graphene-topological insulator van der Waals heterostructures at room temperature. Nat. Comm. *11*, 3657. https://doi.org/10.1038/s41467-020-17481-1.
18. Fabian, J., and Sarma, S.D. (1999). Spin relaxation of conduction electrons. J. Vac. Sci. Technol. B *17*, 1708-1715. https://doi.org/10.1116/1.590813.
19. Averkiev, N.S., and Golub, L.E. (1999). Giant spin relaxation anisotropy in zinc-blende heterostructures. Phys. Rev. B *60*, 15582-15584. https://doi.org/10.1103/PhysRevB.60.15582.
20. Bernevig, B.A., Orenstein, J., and Zhang, S.-C. (2006). Exact SU(2) Symmetry and Persistent Spin Helix in a Spin-Orbit Coupled System. Phys. Rev. Lett. *97*, 236601. https://doi.org/10.1103/PhysRevLett.97.236601.
21. Koralek, J.D., Weber, C.P., Orenstein, J., Bernevig, B.A., Zhang, S.-C., Mack, S., and Awschalom, D.D. (2009). Emergence of the persistent spin helix in semiconductor quantum wells. Nature *458*, 610-613. https://doi.org/10.1038/nature07871.
22. Balocchi, A., Duong, Q.H., Renucci, P., Liu, B.L., Fontaine, C., Amand, T., Lagarde, D., and Marie, X. (2011). Full Electrical Control of the Electron Spin Relaxation in GaAs Quantum Wells. Phys. Rev. Lett. *107*, 136604. https://doi.org/10.1103/PhysRevLett.107.136604.
23. Schliemann, J. (2017). Colloquium: Persistent spin textures in semiconductor nanostructures. Rev. Mod. Phys. *89*, 011001. https://doi.org/10.1103/RevModPhys.89.011001.
24. Tao, L.L., and Tsymbal, E.Y. (2018). Persistent spin texture enforced by symmetry. Nat. Comm. *9*, 2763. https://doi.org/10.1038/s41467-018-

05137-0.

25. Ishizaka, K., Bahramy, M.S., Murakawa, H., Sakano, M., Shimojima, T., Sonobe, T., Koizumi, K., Shin, S., Miyahara, H., Kimura, A., et al. (2011). Giant Rashba-type spin splitting in bulk BiTeI. Nat. Mater. *10*, 521-526. https://doi.org/10.1038/nmat3051.
26. Di Sante, D., Barone, P., Bertacco, R., and Picozzi, S. (2013). Electric Control of the Giant Rashba Effect in Bulk GeTe. Adv. Mater. *25*, 509-513. https://doi.org/10.1002/adma.201203199.
27. Dresselhaus, G. (1955). Spin-Orbit Coupling Effects in Zinc Blende Structures. Phys. Rev. *100*, 580-586. https://doi.org/10.1103/PhysRev.100.580.
28. Hirayama, M., Okugawa, R., Ishibashi, S., Murakami, S., and Miyake, T. (2015). Weyl Node and Spin Texture in Trigonal Tellurium and Selenium. Phys. Rev. Lett. *114*, 206401. https://doi.org/10.1103/PhysRevLett.114.206401.
29. Sakano, M., Hirayama, M., Takahashi, T., Akebi, S., Nakayama, M., Kuroda, K., Taguchi, K., Yoshikawa, T., Miyamoto, K., Okuda, T., et al. (2020). Radial Spin Texture in Elemental Tellurium with Chiral Crystal Structure. Phys. Rev. Lett. *124*, 136404. https://doi.org/10.1103/PhysRevLett.124.136404.
30. Gatti, G., Gosálbez-Martínez, D., Tsirkin, S.S., Fanciulli, M., Puppin, M., Polishchuk, S., Moser, S., Testa, L., Martino, E., Roth, S., et al. (2020). Radial Spin Texture of the Weyl Fermions in Chiral Tellurium. Phys. Rev. Lett. *125*, 216402. https://doi.org/10.1103/PhysRevLett.125.216402.
31. Zhao, H.J., Nakamura, H., Arras, R., Paillard, C., Chen, P., Gosteau, J., Li, X., Yang, Y., and Bellaiche, L. (2020). Purely Cubic Spin Splittings with Persistent Spin Textures. Phys. Rev. Lett. *125*, 216405. https://doi.org/10.1103/PhysRevLett.125.216405.
32. Mera Acosta, C., Yuan, L., Dalpian, G.M., and Zunger, A. (2021). Different shapes of spin textures as a journey through the Brillouin zone. Phys. Rev. B *104*, 104408. https://doi.org/10.1103/PhysRevB.104.104408.
33. Lin, M., Robredo, I., Schröter, N.B.M., Felser, C., Vergniory, M.G., and Bradlyn, B. (2022). Spin-momentum locking from topological quantum chemistry: Applications to multifold fermions. Phys. Rev. B *106*, 245101. https://doi.org/10.1103/PhysRevB.106.245101.
34. Tan, W., Jiang, X., Li, Y., Wu, X., Wang, J., and Huang, B. (2022). A Unified Understanding of Diverse Spin Textures of Kramers–Weyl Fermions in Nonmagnetic Chiral Crystals. Adv. Func. Mater. *32*, 2208023. https://doi.org/10.1002/adfm.202208023.
35. Liu, Y., Li, J., Liu, P., and Liu, Q. (2024). Unconventional spin textures emerging from a universal symmetry theory of spin-momentum locking. npj Quantum Mater. *9*, 69. https://doi.org/10.1038/s41535-024-00682-y.
36. Gosálbez-Martínez, D., Crepaldi, A., and Yazyev, O.V. (2023). Diversity of radial spin textures in chiral materials. Phys. Rev. B *108*, L201114.

https://doi.org/10.1103/PhysRevB.108.L201114.
37. Lin, Z., Wang, C., Xu, Y., and Duan, W. (2020). Hidden physical effects in noncentrosymmetric crystals. Phys. Rev. B *102*, 165143. https://doi.org/10.1103/PhysRevB.102.165143.
38. Yuan, L.-D., Wang, Z., Luo, J.-W., Rashba, E.I., and Zunger, A. (2020). Giant momentum-dependent spin splitting in centrosymmetric low-Z antiferromagnets. Phys. Rev. B *102*, 014422. https://doi.org/10.1103/PhysRevB.102.014422.
39. Yuan, L.-D., Wang, Z., Luo, J.-W., and Zunger, A. (2021). Prediction of low-Z collinear and noncollinear antiferromagnetic compounds having momentum-dependent spin splitting even without spin-orbit coupling. Phys. Rev. Mater. *5*, 014409. https://doi.org/10.1103/PhysRevMaterials.5.014409.
40. Liu, K., Luo, W., Ji, J., Barone, P., Picozzi, S., and Xiang, H. (2019). Band splitting with vanishing spin polarizations in noncentrosymmetric crystals. Nat. Comm. *10*, 5144. https://doi.org/10.1038/s41467-019-13197-z.
41. Ji, J., Lou, F., Yu, R., Feng, J.S., and Xiang, H.J. (2022). Symmetry-protected full-space persistent spin texture in two-dimensional materials. Phys. Rev. B *105*, L041404. https://doi.org/10.1103/PhysRevB.105.L041404.
42. Reehuis, M., Krimmel, A., Büttgen, N., Loidl, A., and Prokofiev, A. (2003). Crystallographic and magnetic structure of ZnV2O4. Eur. Phys. J. B *35*, 311-316. https://doi.org/10.1140/EPJB/E2003-00282-4.
43. Železný, J., Zhang, Y., Felser, C., and Yan, B. (2017). Spin-Polarized Current in Noncollinear Antiferromagnets. Phys. Rev. Lett. *119*, 187204. https://doi.org/10.1103/PhysRevLett.119.187204.
44. Lee, S.H., Qian, Y., and Yang, B.-J. (2024). Fermi Surface Spin Texture and Topological Superconductivity in Spin-Orbit Free Noncollinear Antiferromagnets. Phys. Rev. Lett. *132*, 196602. https://doi.org/10.1103/PhysRevLett.132.196602.
45. Tomeno, I., Fuke, H.N., Iwasaki, H., Sahashi, M., and Tsunoda, Y. (1999). Magnetic neutron scattering study of ordered Mn3Ir. J. Appl. Phys. *86*, 3853-3856. https://doi.org/10.1063/1.371298.
46. Guo, S.-D., Feng, X.-K., Huang, D., Chen, S., Wang, G., and Ang, Y.S. (2023). Intrinsic persistent spin texture in two-dimensional T-XY (X, Y=P, As, Sb, Bi; X $\neq$ Y). Phys. Rev. B *108*, 075421. https://doi.org/10.1103/PhysRevB.108.075421.
47. Chang, G., Wieder, B.J., Schindler, F., Sanchez, D.S., Belopolski, I., Huang, S.-M., Singh, B., Wu, D., Chang, T.-R., Neupert, T., et al. (2018). Topological quantum properties of chiral crystals. Nat. Mater. *17*, 978-985. https://doi.org/10.1038/s41563-018-0169-3.
48. Wang, J., Sui, X., Gao, S., Duan, W., Liu, F., and Huang, B. (2019). Anomalous Dirac Plasmons in 1D Topological Electrides. Phys. Rev. Lett.

*123*, 206402. https://doi.org/10.1103/PhysRevLett.123.206402.
49. Bradlyn, B., Elcoro, L., Cano, J., Vergniory, M.G., Wang, Z., Felser, C., Aroyo, M.I., and Bernevig, B.A. (2017). Topological quantum chemistry. Nature *547*, 298-305. https://doi.org/10.1038/nature23268.
50. Vergniory, M.G., Elcoro, L., Felser, C., Regnault, N., Bernevig, B.A., and Wang, Z. (2019). A complete catalogue of high-quality topological materials. Nature *566*, 480-485. https://doi.org/10.1038/s41586-019-0954-4.
51. Vergniory, M.G., Wieder, B.J., Elcoro, L., Parkin, S.S.P., Felser, C., Bernevig, B.A., and Regnault, N. (2022). All topological bands of all nonmagnetic stoichiometric materials. Science *376*, abg9094. https://doi.org/10.1126/science.abg9094.
52. Yang, B.-J., Bojesen, T.A., Morimoto, T., and Furusaki, A. (2017). Topological semimetals protected by off-centered symmetries in nonsymmorphic crystals. Phys. Rev. B *95*, 075135. https://doi.org/10.1103/PhysRevB.95.075135.
53. Bradlyn, B., Cano, J., Wang, Z., Vergniory, M.G., Felser, C., Cava, R.J., and Bernevig, B.A. (2016). Beyond Dirac and Weyl fermions: Unconventional quasiparticles in conventional crystals. Science *353*, aaf5037. https://doi.org/10.1126/science.aaf5037.
54. Koyama, S., and Rondinelli, J.M. (2023). Accidental persistent spin textures in the proustite mineral family. Phys. Rev. B *107*, 035154. https://doi.org/10.1103/PhysRevB.107.035154.
55. Lee, H., Choi, B., and Lee, H.-W. (2022). Orientational dependence of intrinsic orbital and spin Hall effects in hcp structure materials. Phys. Rev. B *105*, 035142. https://doi.org/10.1103/PhysRevB.105.035142.
56. Yang, Q., Xiao, J., Robredo, I., Vergniory, M.G., Yan, B., and Felser, C. (2023). Monopole-like orbital-momentum locking and the induced orbital transport in topological chiral semimetals. Proc. Natl. Acad. Sci. U.S.A. *120*, e2305541120. https://doi.org/10.1073/pnas.2305541120.
57. Brinkman, S.S., Tan, X.L., Brekke, B., Mathisen, A.C., Finnseth, Ø., Schenk, R.J., Hagiwara, K., Huang, M.-J., Buck, J., Kalläne, M., et al. (2024). Chirality-Driven Orbital Angular Momentum and Circular Dichroism in CoSi. Phys. Rev. Lett. *132*, 196402. https://doi.org/10.1103/PhysRevLett.132.196402.
58. Kresse, G., and Furthmüller, J. (1996). Efficiency of ab-initio total energy calculations for metals and semiconductors using a plane-wave basis set. Comput. Mater. Sci. *6*, 15-50. https://doi.org/10.1016/0927-0256(96)00008-0.
59. Kresse, G., and Furthmüller, J. (1996). Efficient iterative schemes for ab initio total-energy calculations using a plane-wave basis set. Phys. Rev. B *54*, 11169-11186. https://doi.org/10.1103/PhysRevB.54.11169.
60. Perdew, J.P., Burke, K., and Ernzerhof, M. (1996). Generalized Gradient Approximation Made Simple. Phys. Rev. Lett. *77*, 3865-3868. https://doi.org/10.1103/PhysRevLett.77.3865.

61. Gallego, S.V., Perez-Mato, J.M., Elcoro, L., Tasci, E.S., Hanson, R.M., Momma, K., Aroyo, M.I., and Madariaga, G. (2016). MAGNDATA: towards a database of magnetic structures. I. The commensurate case. J. Appl. Cryst. *49*, 1750-1776. https://doi.org/10.1107/S1600576716012863.
62. Aroyo, M.I., Perez-Mato, J.M., Capillas, C., Kroumova, E., Ivantchev, S., Madariaga, G., Kirov, A., and Wondratschek, H. (2006). Bilbao Crystallographic Server: I. Databases and crystallographic computing programs. Z. Kristallogr. - Cryst. Mater. *221*, 15-27. https://doi.org/10.1524/zkri.2006.221.1.15.
63. Liu, P., Li, J., Han, J., Wan, X., and Liu, Q. (2022). Spin-Group Symmetry in Magnetic Materials with Negligible Spin-Orbit Coupling. Phys. Rev. X *12*, 021016. https://doi.org/10.1103/PhysRevX.12.021016.